\documentclass[paper]{aa}
\usepackage{graphicx}
\usepackage{txfonts}
\def\xmm{XMM-{\it Newton}}

\def\sw{{\it Swift}}
\def\frm{{\it Fermi}}
\def\igr{{\it Integral}}

\def\bi{\textsc{BatImager}~}

\def\2e{$\times$ 10$^{-2}$}
\def\3e{$\times$ 10$^{-3}$}
\def\*{$^{\star}$}
\def\1u{1FGL~J0137.8+5814}
\def\2u{1FGL~J2056.7+4938}

\begin{document}

\title{The populations of hard X- and $\gamma$-ray sources: \\
a correlation study and new possible identifications}
\author{A.~Maselli\inst{1}, G.~Cusumano\inst{1}, E.~Massaro\inst{2},
  A.~Segreto\inst{1}, V.~La~Parola\inst{1}, A.~Tramacere\inst{3},
  I.~Donnarumma\inst{4}.}

\institute{INAF, Istituto di Astrofisica Spaziale e Fisica Cosmica di Palermo,
  Via U. La Malfa 153, I-90146 Palermo, Italy \and Dipartimento di Fisica,
  Universit\`a La Sapienza, Piazzale A. Moro 2, I-00185 Roma, Italy \and
  INTEGRAL Science Data Centre, CH-1290 Versoix, Switzerland \and INAF IASF
  Roma, via Fosso del Cavaliere 100, I-00133 Roma, Italy }
       
\offprints{maselli@ifc.inaf.it} 

\date{Received 18 March 2011 / accepted 09 April 2011}

\titlerunning{The populations of hard X- and $\gamma$-ray sources}

\authorrunning{A.~Maselli et al.}

\abstract
{}
{
  We present the results of our analysis devoted to the research of sources
  emitting in the energy bands surveyed by both the \sw-BAT and the \frm-LAT
  telescopes.
}
{
  We cross-correlate the \frm-LAT 1-year point source catalogue (1FGL) of
  gamma-ray sources and the second Palermo BAT catalogue (2PBC) of hard X-ray
  sources, establishing a correspondence between sources when their error
  boxes overlap. We also extract the significance value in the BAT 15--150
  keV map, obtained using a dedicated software for the reduction of BAT data,
  in the direction of the 1FGL sources and take into account those above the
  significance threshold $\sigma = 3$.
}
{
   We obtain a sample of common sources emitting in both the hard X- and the
  $\gamma$-ray energy bands and evaluate its content in galactic and
  extragalactic objects. We assess the fraction of unidentified sources and
  describe in greater detail the properties of two of them, \1u~and \2u,
  supporting their classification as blazars after the analysis of their
  broad-band spectral energy distribution. We discuss the blazar content of
  the collected 1FGL-2PBC sources: we build its redshift distibution and
  compare it with that of the whole blazar population as reported in the
  second edition of the BZCAT blazar catalogue.
}
{}


\maketitle


\section{Introduction}
\label{sect:1}

The present generation of space observatories for high energy astrophysics is
characterised by large area and wide field instrumentation.
This is the case of the Burst Alert Telescope (BAT, \cite{barthelmy2005}) and
Large Area Telescope (LAT, \cite{atwood2009}) onboard \sw~and \frm-GST,
respectively.
One of the main throughputs of these instruments is the discovery of a thousand
of hard X- and $\gamma$-ray sources and the possibility of performing
investigations on the population properties and evolution much more accurate
than in the past.

\begin{figure*}[hbtp]
\begin{center}
\includegraphics[width=18cm]{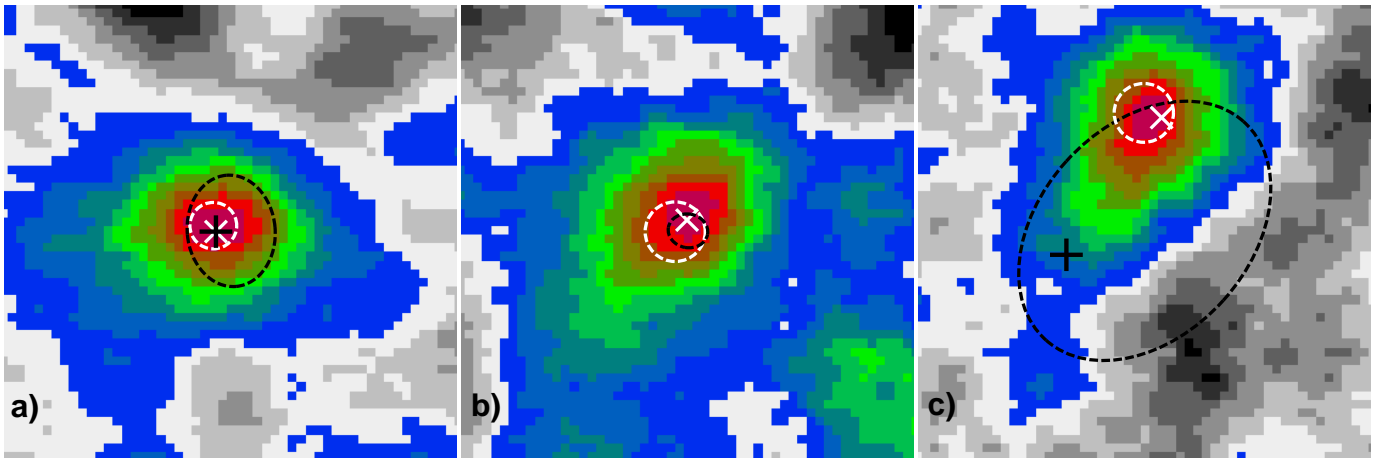}
\end{center}
\caption{Details of the BAT significance map at the position of a)
  1FGL~J1103.7$-$2329 (left panel), b) 1FGL~J2056.7+4938 (middle panel)
  and c) 1FGL~J0238.3$-$6132 (right panel). A dashed line represents the
  position of the 1FGL (black ellipse) and the 2PBC (white circle)
  sources with its uncertainty. For each source, the position of the
  associated counterpart is indicated by a cross of the corresponding
  colour. Each square map is $\sim 60' \times 60'$; the colour scale
  is optimised for each hard X-ray source.}
\label{figura01}
\end{figure*}

The first catalogue of $\gamma$-ray point sources (1FGL, \cite{abdo2010a}),
based on data obtained in the 11 months after the beginning of scientific
operation (2008, August 4) contains 1451 entries, and a fraction of about
40~\% of them does not have a reliable counterpart.
A rather similar situation is also occurring with the hard X-ray sources
detected by BAT: the second Palermo BAT catalogue (\cite{cusumano2010b},
hereafter 2PBC), that covers the observation period from November 2004 to May
2009 and is the richest one in this energy range, also includes 177 not
associated sources, over a total number of 1256.
A multifrequency approach based on the cross-correlation of catalogues in
different energy bands from the radio to the $\gamma$ rays can be very usesul,
if not essential, to unravel the nature of unidentified sources.

In this paper we use the 1FGL and the 2PBC catalogues to compare the
populations of hard X- and $\gamma$-ray sources and to search for
possible correspondences between them.
The paper is organised as follows: we report the details of the
cross-correlation between the two catalogues in Section~\ref{sect:2}
and of the cross-correlation of the 1FGL catalogue with the BAT
15--150 keV all-sky significance map in Section~\ref{sect:3}.
We discuss our results in Section~\ref{sect:4} giving a particular
emphasis on the blazar content of the collected group of sources.
In Section~\ref{sect:5} we describe the properties of two unidentified
objects in the 1FGL catalogue, at low Galactic latitude, with a
significant emission in the hard X-ray band: we collect all the
available data in the literature and analyse in detail their broad
band spectral energy distribution (SED) to support their
classification as blazars.
The main results of our analysis are summarised in
Section~\ref{sect:6}.

\section{Correspondences between 1FGL and 2PBC sources}
\label{sect:2}

The 2PBC catalogue has been obtained from the reduction of the BAT data
collected in 54 months since the launch of the \sw~mission using the dedicated
software \bi (\cite{segreto2010}).
It contains 1256 hard X-ray sources detected at a significance level higher
than $\sigma_T = 4.8$; their coordinates are given with a 95\% confidence
level radius $r_{BAT}$.
A counterpart was associated to 1079 sources ($\sim 86\%$); for 26 of them a
double association was found, and in two cases three possible counterparts
were proposed; sources without any associated counterpart are 177.

We match the 2PBC catalogue with the 1FGL catalogue to search for sources
emitting in both the energy bands surveyed by the \sw-BAT and the \frm-LAT
telescopes.
The positions of the 1451 1FGL $\gamma$-ray sources, determined by means of a
maximum likelihood algorithm, are given with an uncertainty ellipse
corresponding to the 95\% probability of locating the source; 58 objects are
reported without any positional uncertainty.
We adopt, when available, the mean value of the two axes of the 95\%
confidence ellipse as the radius $r_{LAT}$ of the error circle for a
1FGL~source.
We calculate the angular distance $d$ among the centroids of 1FGL and 2PBC
sources and establish a correspondence when the two error circles overlap,
adopting the condition $d \leq (r_{BAT} + r_{LAT})$: applying this criterion
we obtain 77 correspondences.
We define a simple parameter to estimate the quality of the correspondence and
use the expression $Q = d / r_{BL}$, where $r_{BL}$ is the higher value
between $r_{BAT}$ and $r_{LAT}$.
The parameter $Q < 1$ implies that the wider error circle includes both
centroids: this value occurs for 64 correspondences, while in four cases we
find $Q > 1.4$.
We consider the possibility that the choice of the mean value of the axes of
the 95\% uncertainty ellipse, adopted to estimate the 1FGL uncertainty region,
may be not firmly reliable.
This aspect raises particularly in cases of high eccentricity values, in which
the orientation of the 1FGL ellipse with respect to the 2PBC circle must be
taken into account.
In order to gain confidence about the established correspondence between
sources we compute the axial ratio of the uncertainty ellipse and verify in
the sky-map the 25 cases ($\sim 1/3$) in which we find values higher than 1.2.
We find that in all but four cases not only the centroid of the 2PBC source,
but also the associated counterpart, is inside the 1FGL ellipse.

\setcounter{table}{4}                                                         
\begin{table*}[bhtp]
  \caption{A statistical description of the blazar content of sources
    showing a simultaneous emission both in the BAT and in the LAT energy
    ranges. Left column: all the blazars classified in the BZCAT. Middle
    column: blazars obtained by the cross-correlation of 1FGL with the 2PBC
    catalogues. Right column: blazars with a significance higher than
    $\sigma_T^{\star} = 3$. The mean redshift value is reported for each
    subsample. Values in brackets refer to firm redshift estimates, or
    different from zero.}
\label{tabella01}
\begin{center}
\begin{tabular}{ccc|cc|cc}
type  &      BZCAT     & $\langle z \rangle$ & 1FGL-2PBC & $\langle z_1 \rangle$ & $\sigma > \sigma_T^{\star}$ & $\langle z_{2} \rangle$ \\
\hline
BZB   & 1164  (531)    &   $0.33 \pm 0.01$   &  16 (14)  &   $0.17 \pm 0.06$     &    24 (19)        &    $0.21 \pm 0.05$      \\
BZQ   & 1660 (1638)    &   $1.40 \pm 0.02$   &      27   &   $1.39 \pm 0.17$     &         41        &    $1.23 \pm 0.12$      \\
BZU   &  262  (215)    &   $0.42 \pm 0.04$   &       7   &   $0.32 \pm 0.14$     &          8        &    $0.32 \pm 0.12$      \\
\hline                                                                                               
total & 3086 (2384)    &                     &  50 (48)  &                       &    73 (68)        &                         \\            
\end{tabular}
\end{center}
\end{table*}

Further correspondences have been established for 9 of the 58 1FGL~sources
with no positional uncertainty as the position of the 1FGL source is within
the error circle of the 2PBC source.
The remaining 49 cases are all unambiguously discarded as no 2PBC source is
found in the proximity of these 1FGL sources.
The final list includes 86 1FGL-2PBC correspondences for a total of 84 1FGL
sources: two of them have in fact a possible correspondence with two different
2PBC sources.
Moreover, we find also a correspondence in which two possible
counterparts, both high mass X-ray binaries, have been
associated to the same 2PBC source in the Small Magellanic Cloud.
The list of correspondences has been splitted in two parts reported in
Table~\ref{tabella02} and Table~\ref{tabella03} according to the Galactic
latitude, with 63 correspondences at $|b|>10^{\circ}$ and 23 at
$|b|<10^{\circ}$.
For each correspondence we report the 1FGL and 2PBC identifiers, the
associated counterparts and their classification.
For a few AGNs the counterpart, missing in the 1FGL catalogue, has
been found in the first catalogue of active galactic nuclei detected
by the Fermi Large Area Telescope (\cite{abdo2010b}): these cases have
been marked with an asterisk ($^{\ast}$) in Table~\ref{tabella02}.
The value of the $Q$ parameter and the agreement between the corresponding
counterparts is reported in the last two columns.
We mark with a colon the $Q$ values of the four correspondences for
which the 2PBC counterparts are outside the 1FGL uncertainty ellipse.
We find also two cases (marked with ``c'') having a possibility of
multiple association: this confusion is due to crowded fields, one of
them close to the Galactic centre direction.
In 62 cases (marked with ``y'') the counterpart is the same, while in 12 cases
(marked with ``n'') the counterparts associated with the high-energy sources
are different: in these cases the correspondence is supposed to be due to
chance.
We find also 13 correspondences for which both sources, or just one of them,
lack the associated counterpart.
Assuming that the 1FGL-2PBC correspondence is the result of high-energy
emission from a single source, this could be considered a hint toward the
association of a counterpart to some of these 1FGL sources; all these
correspondences, with only one exception, are characterised by a $Q$ parameter
lower than unity.

We plot in Fig.~\ref{figura01} the details of the BAT 15--150 keV significance
map for some of these correspondences.
In the left panel of Fig.~\ref{figura01} we report the case in which the 1FGL
and the 2PBC sources correspond to the same object, the BL~Lac 1H~1100$-$230.
In the middle panel we report the case of the source \2u that has not been
identified in the 1FGL catalogue: the correspondence with the 2PBC source
suggests the X-ray source RXJ2056.6+4940 as possible counterpart.
The properties of this source have been investigated in greater detail and
discussed in Section~\ref{sub02}.
Finally, in the right panel of Fig.~\ref{figura01} we report the case
of a correspondence probably due to chance: two close objects, the
galaxy IRAS F02374$-$6130 and the flat spectrum radio quasar
PKS~0235$-$618, are responsible for the hard X- and the $\gamma$-ray
emission, respectively.

We compare our results with those reported in Abdo et~al.~(2010b, their
Table~7) and find that all the 50 sources provided in their list are included
in Table~\ref{tabella02}, with the exception of 1FGL~J1938.2$-$3957.
Hard X-ray emission in this direction was revealed by \igr~and reported in the
fourth IBIS catalogue (\cite{bird2010}): the counterpart associated to this 1FGL
source, PKS~1933$-$400, is also reported in the BZCAT as a blazar with uncertain
classification.
The value $\sigma \sim 3.5$ that we find in the BAT 15--150 keV significance
map at the position of PKS~1933$-$400 is compatible with hard X-ray emission
from this source, but it is lower than $\sigma_T$.

\section{Correspondences of 1FGL sources with the 54-month hard X-ray maps}
\label{sect:3}

The 2PBC catalogue includes sources with a significance threshold
$\sigma_T=4.8$.
Reasonably, a remarkable number of fainter objects are imaged in the 54-month
BAT all-sky maps at a significance level lower than $\sigma_T$.
In the effort of increasing the list of objects emitting both in the BAT and
in the LAT energy ranges we consider the adoption of a lower significance
threshold.
Taking into account the results obtained by Maselli et~al.~(2010) in the
cross-correlation of the BZCAT Blazar Catalogue (\cite{massaro2009}) with the
39-month 15--150 keV BAT map (\cite{cusumano2010a}) we adopt
$\sigma_T^{\star}=3$.
From their analysis, carried out at $|b|>10^{\circ}$, Maselli et~al.~(2010)
found that the adoption of $\sigma_T^{\star}$ follows in a fraction of
$\sim$~3\% of related spurious associations; this value is presumably
underestimated at $|b|<10^{\circ}$ due to the higher density of sources.

A total of 1043 and 408 objects are found in the 1FGL catalogue at high
($|b|>10^{\circ}$) and low Galactic latitude, respectively.
We extract the significance value in the 54-month 15-150 keV BAT all-sky map
at their positions and find $\sigma \geq \sigma_T^{\star}$ for 80 objects at
$|b|>10^{\circ}$ and 49 objects at $|b|<10^{\circ}$.
In a few of these objects the $\sigma$ value may be strongly biased by some
close, very bright hard X-ray sources.
Following the same criterion adopted in Maselli et~al.~(2010), we exclude the
1FGL sources whose position is found within $36'$ from these bright BAT
sources and no overlap, even marginal, is found between the corresponding
error regions.
After this screening we obtain a sample of 75 ($|b|>10^{\circ}$) and 29
($|b|<10^{\circ}$) 1FGL sources.
All the sources listed in Table~\ref{tabella02} and Table~\ref{tabella03} are
included in this sample with nine exceptions, all of them at high Galactic
latitude.
They concern 1FGL sources for which the 95\% confidence ellipse is
particularly wide and the 2PBC source is at a considerable distance from the
centre of the ellipse.
The final list of additional sources with respect to those obtained
from the cross-correlation between the 1FGL and 2PBC catalogues has
been splitted according to the Galactic latitude and reported in
Table~\ref{tabella04} and Table~\ref{tabella05}.

\section{Properties of the resulting samples of sources}
\label{sect:4}

\begin{figure*}
\begin{center}
\includegraphics[width=6.9cm,angle=-90]{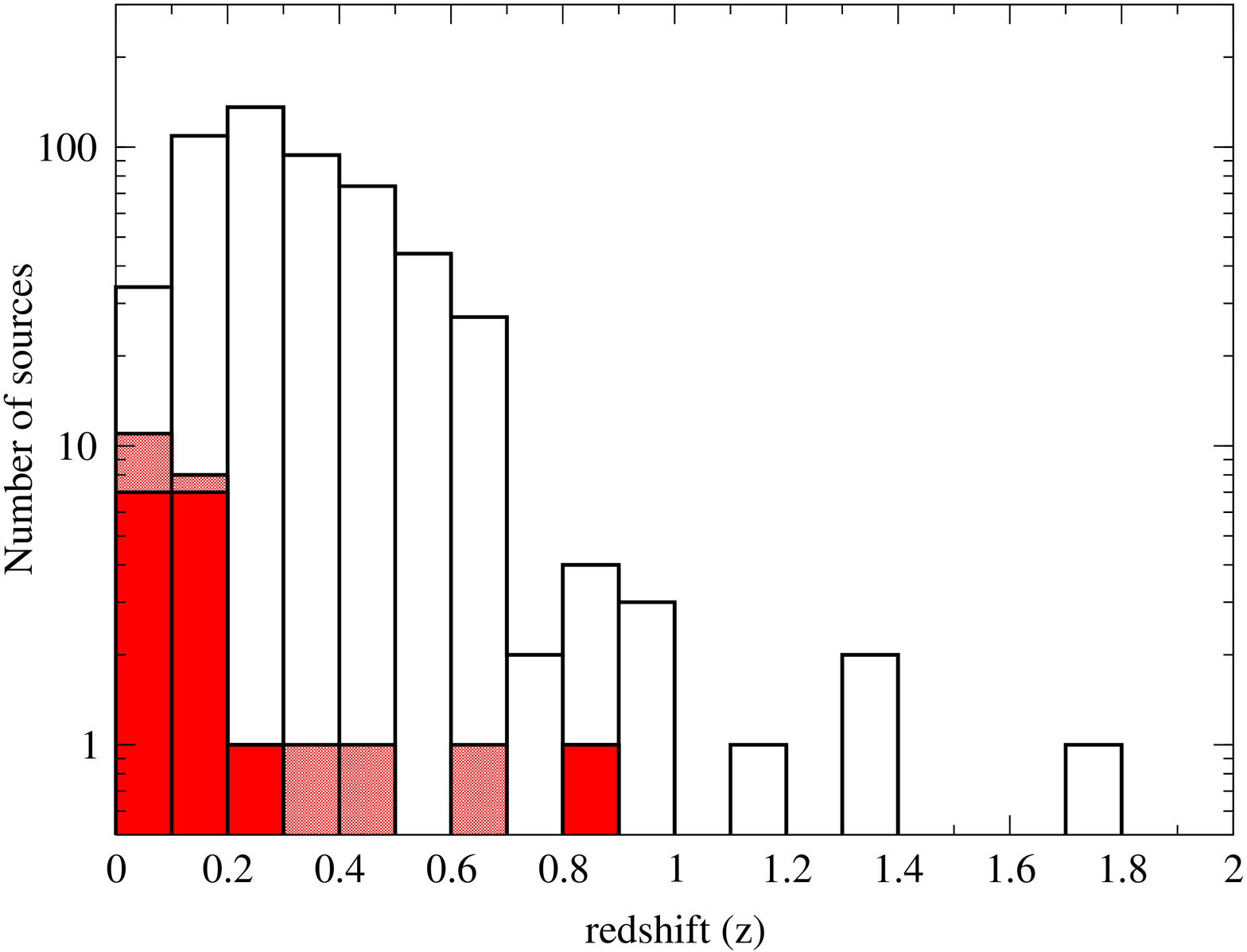} \includegraphics[width=6.9cm,angle=-90]{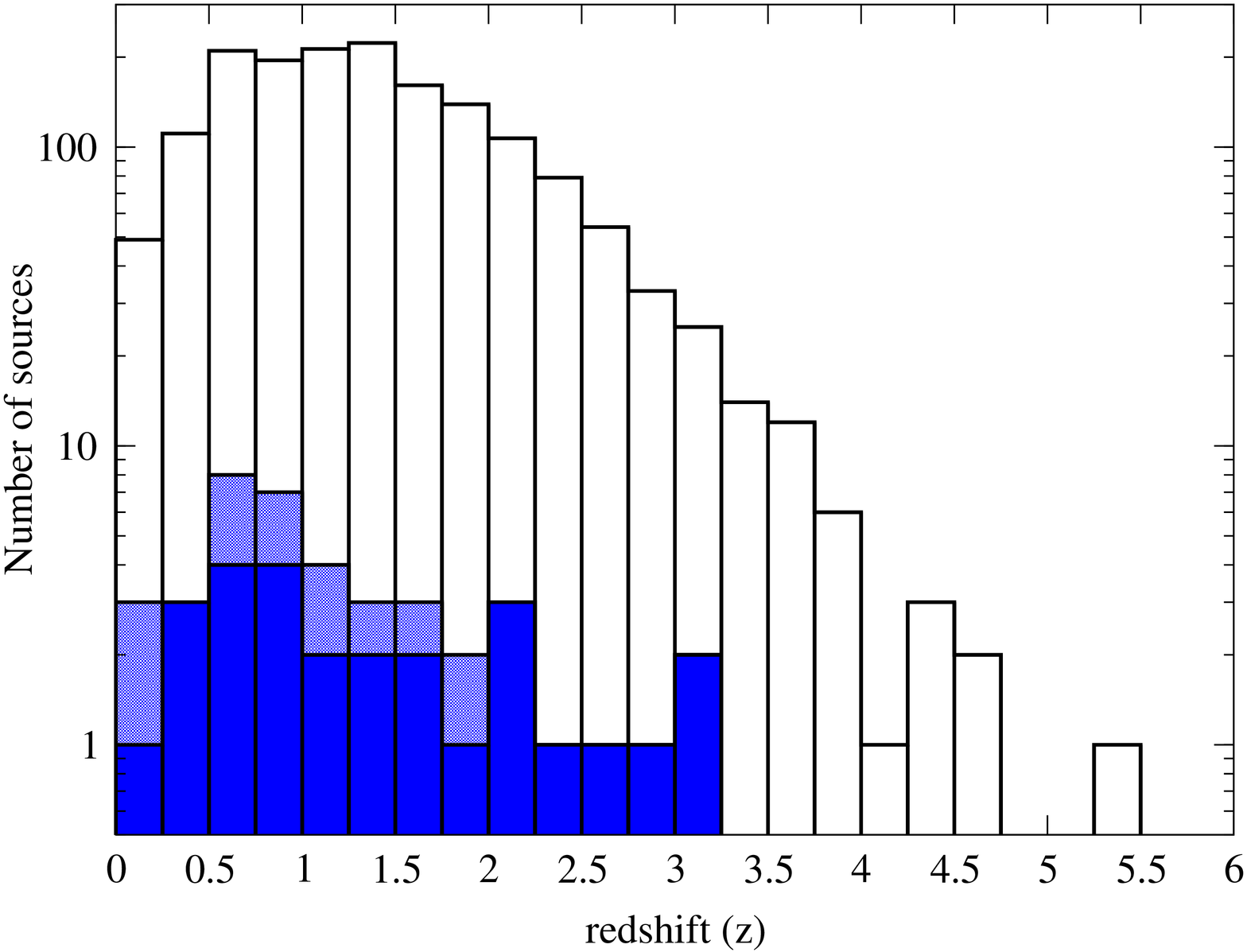} 
\end{center}
\caption{Histograms of the redshift distributions for BL~Lac objects (left
  panel) and flat spectrum radio quasars (right panel). The contribution of
  1FGL sources with a significance down to $\sigma_T^{\star}=3$ in the BAT
  15--150 keV significance map (filled columns) is compared with that of the
  whole corresponding population reported in the BZCAT (empty columns). The
  contribution of sources included in the 2PBC catalogue is emphasised by a
  thicker colour.}
\label{figura04}
\end{figure*}

The firm correspondences that we find cross-correlating the 1FGL and 2PBC
catalogues and verifying the agreement between the identified counterparts
lead to 62 sources.
This number raises to 104 ($\sim 7\%$ of all the 1FGL objects) considering all
the sources with a significance down to $\sigma_T^{\star} = 3$.
Therefore, the number of sources emitting in both the energy ranges covered by
the BAT and the LAT instruments aboard \sw~and \frm, respectively, is small.
This is a clear indication that the emission in the hard X- and in the
$\gamma$-ray sky is dominated by sources with a different nature.
This result is not unexpected: among extragalactic sources, the main
contribution in the 1FGL catalogue is given by blazars, while in the 2PBC
catalogue it is given by Seyfert galaxies.
Considering the 1FGL extragalactic sources provided with an association, the
blazar contribution is given by 295 BL~Lacs and 274 FSRQs: active galaxies
with uncertain classification are 92, while non-blazar active galaxies are
only 28.
Conversely, in the 2PBC catalogue there are 307 Seyfert~1 and 165 Seyfert~2
galaxies. 
Even if considerable, the blazar contribution (97 objects) is less relevant.

The group of 104 objects that we have collected is made up of 83 extragalactic
and 15 galactic sources; 6 objects are unidentified.
The largest part of extragalactic sources is given by blazars with only a very
few exceptions: the Seyfert 1.2 galaxy ESO~323$-$77, the FR~II radiogalaxy
3C~111, the flat spectrum radio source PKS~0336$-$177 and the starburst galaxies
M~82 and NGC~4945.  
At low Galactic latitude we find 6 pulsars (including Crab and Vela), 4 high
mass X-ray binaries (including Cyg~X-3) and a low mass X-ray binary, the
supernova remnant Cas~A, two cataclismic variable stars and the peculiar
object Eta~Carinae.
We briefly report on the results obtained considering the lower
significance threshold value $\sigma = 2$: we find further 74 sources,
the largest fraction of which are blazars (16~BZB, 18~BZQ and 9 BZU),
followed by other AGNs (among which the well known FR~I radiogalaxy
M87) the starburst galaxy NGC~253 and a few pulsars; 14 sources are
unidentified.
We remark that, according to Maselli et~al.~(2010), the number of spurious
correspondences for such a low significance threshold is $\sim 20\%$.

We focus on the blazar content of the 1FGL-2PBC sources and investigate about
their distance.
We follow the classification of the second edition of the BZCAT catalogue
(\cite{massaro2010}) which includes 1164 BL~Lac objects (indicated in the
following with the suffix ``B''), 1660 flat spectrum radio quasars (suffix
``Q'') and 262 blazars with uncertain classification (suffix ``U'').
The number of BL~Lacs without redshift estimate (541) is considerable:
moreover, in our computation we consider only firm estimates, for a total of
531 BL~Lacs.
We compute the mean redshift value of the different classes of 1FGL-2PBC
blazars: the value $\langle z_{1} \rangle$ refers to the group of blazars
obtained from the cross-correlation of the 1FGL and 2PBC catalogues while
$\langle z_{2} \rangle$ refers to the larger group obtained adopting the lower
threshold value $\sigma_T^{\star} = 3$ in the BAT 15--150 significance map.
We note that 5 among the 24 BL~Lacs detected above $\sigma_T^{\star}$ do not
have any redshift estimate, two of which are included in the 2PBC catalogue.
We compare these results with the values which characterise the blazar
subclasses as a whole, computed considering the totality of sources classified
in the BZCAT: all these values are reported in Table~\ref{tabella01}.
Moreover, we plot in Fig.~\ref{figura04} the histograms of the redshift
distributions for BL~Lacs (left panel) and flat spectrum radio quasars (right
panel) for the considered groups of sources.

Our results show that the subsample of blazars emitting both in the BAT and in
the LAT energy bands is made of sources relatively closer than the average to
the observer.
In fact, $\langle z_{2} \rangle$ is lower than the mean redshift value
computed for all the different blazar subclasses, as reported in
Table~\ref{tabella01}.
The analysis of the plot in the left panel of Fig.~\ref{figura04} shows that
the largest part of BL~Lac objects coming from the cross-correlation of 1FGL
and 2PBC catalogues has a redshift $z_B < 0.2$, while the modal value of the
BL~Lac distribution is in the range $0.2 < z_B < 0.3$.
The addition of sources with significance down to $\sigma_T^{\star} = 3$
confirms this result: the largest part of them have $z_{B} < 0.1$, with the
remaining sources more or less equally distributed at higher redshift up to
$z_{B} = 0.7$.
An analogous result is true for flat spectrum radio quasars
(Fig.~\ref{figura04}, right panel): high-energy sources emitting both in the
BAT and in the LAT energy bands have redshift $z_{Q}<3.2$, with a peak in
the range $0.5<z_{Q}<1$.
The addition of sources with significance down to $\sigma_T^{\star}=3$
increments this peak with further 7 sources, and none of them has a redshift
higher than $z_{Q}=2$.
Conversely, the modal value of the distribution of all the FSRQs catalogued
in the BZCAT is at higher values than this peak, in the range $1.25 < z_Q <
1.5$.

\section{New possible associations of sources in the BAT and LAT surveys}
\label{sect:5}

The high-energy emission revealed by the BAT and the LAT telescopes is very
helpful in addressing a more correct classification of already known sources.
Moreover, it can lead to the discovery of new blazars in regions where their
identification is complicated by the belt surrounding the Galactic plane
intercepting the line of sight towards them.
We focus our attention on two 1FGL unidentified sources, \1u and \2u, detected
at low Galactic latitude and included in Table~\ref{tabella03} and
Table~\ref{tabella05}.
For each source we build the spectral energy distribution (SED) by
adding the data obtained from our reduction of \sw~and \xmm~pointed
observations to all the data that we have found in the literature.

\sw-XRT observations, carried out using the most sensitive Photon Counting
readout mode (see \cite{hill2004} for a description of readout modes), are
available for both \1u and \2u.
The highest detected count rate for each source is in any case lower than the
pile-up threshold (0.5~cts~s$^{-1}$).
We reduce the data with the HEASOFT~6.8 package distributed by the NASA High
Energy Astrophysics Archive Research Center (HEASARC).
All the files necessary for the spectral analysis are obtained using the
{\sc xrtpipeline} and the {\sc xrtproducts} tasks.
For each source two \sw~observations, characterised by very similar values of
the count rate, are available: therefore we decide to sum the two event
files using the {\sc xselect} task and the exposure maps to correct for
vignetting, CCD hot and damaged pixels.
We use the {\sc xrtcentroid} task to detect the centroid of the source
and the corresponding positional error; a nearby source-free region is
chosen for the extraction of the background spectrum.
Spectral data of sources are extracted in circular regions surrounding
the centroid, adopting a radius of 20~pixels (1 pixel = $2.36''$) for
the source spectrum and of 50~pixels for the background spectrum.

A pointed observation from \xmm~is available for \1u: we restrict our
analyses of this observation to MOS1 and MOS2, discarding PN data
because the source is located at the edge of the instrument FoV.
We use the standard analysis software SAS~10.0 to extract high level science
products from the ODF files.
The source spectrum is extracted by selecting all the events with PATTERN~$\le
12$ (restricting the patterns to single and doubles) and FLAG~$=0$ within a
circle of $40''$ radius centered on the source.
Similarly, the background is extracted collecting all the counts within an
annulus, centered on the source, with inner and outer radii of $45''$ and
$85''$, respectively.
We combine the instrumental channels in the spectral files to include at
least 20 counts in each new energy bin for the \sw~observation and 25 counts
in the case of \xmm; the spectral analysis is carried out using XSPEC 12.5.1n.

The flux density of both sources in the hard X-ray domain (20~keV) has been
obtained converting their count rate in the 15--30 keV map of the 54-month BAT
survey.
The conversion factor is calculated from the count rate of Crab and
its spectrum used for calibration purposes, as explained in the BAT
calibration status
report\footnote{http://swift.gsfc.nasa.gov/docs/swift/analysis/bat\_digest.html}.

The measurements available for these objects at different frequencies are
mostly non-simultaneous and their number is not sufficient to allow the
possibility of analysing in detail their intrinsic variability at different
epochs.
We evaluate their general emission properties by selecting opportune energy
ranges and, whenever possible, by fitting the corresponding data with
analytical models to estimate some relevant parameters.
The data in the radio band are well fitted by a power-law model
$S(\nu)~\propto~\nu^{-\alpha_r}$ to derive their spectral index $\alpha_r$.
We adopt a log-parabolic model to fit the curvature of energy distribution
of the synchrotron and inverse Compton components that characterise the SED 
of blazars.
This model, expressed by the analytical formula 
\begin{equation} 
\nu F(\nu) = \nu_p\,F(\nu_p)\times 10^{-\beta\,(\mathrm{Log}~\nu/\nu_p)^2}, 
\end{equation} 
provides the peak frequency $\nu_p$ of the component and the parameter $\beta$ 
that describes its curvature at the peak.
This law reproduces curved spectra with a small number of parameters and has 
been verified to fit well the broad band spectra of blazars (\cite{landau1986}, 
see also Massaro et al. 2004a,b for an interpretation in terms of statistical 
acceleration).

We note that the Galactic latitude of these sources is very low.
For this reason the flux density in the optical band may be severely affected
from uncertainties of the Galactic extinction values which are supposed to be
not fully reliable.
As regards the $\gamma$-ray data, derived from the 1FGL catalogue, we consider
the possibility of contamination due not only to the $\gamma$-ray background
but also to the occasional presence of neighbour sources in the field as for
the case of \1u, with a pulsar at an angular distance of $\sim 12'$.

\subsection{\1u}
\label{sub01}

We assume from the 1FGL catalogue the position of \1u (RA=$01^h 37^m 48^s.77$;
Dec=$+58^{\circ} 14' 57''.1$) with an error radius $7'.5$; the Galactic
latitude is $b=-4^{\circ}.07$.

\begin{figure}[htbp]
\begin{center}
\includegraphics[width=8.8cm]{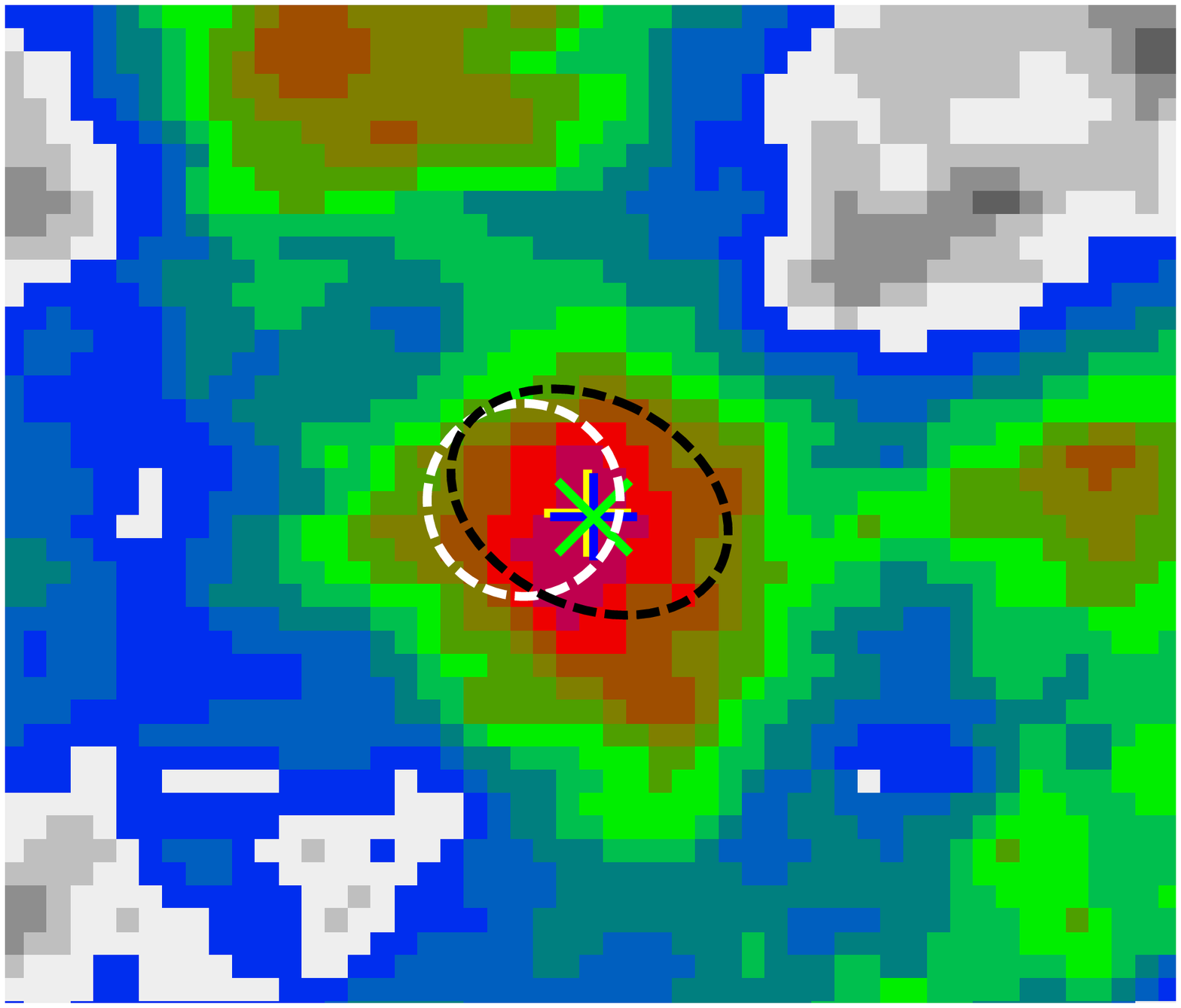}
\end{center}
\caption{The BAT 15--150 keV significance map ($60' \times 60'$)
    in the field of \1u. The positions of the Fermi-LAT and the
    \igr-IBIS detections with the corresponding uncertainty regions
    are plotted with a black and a white dashed line,
    respectively. The positions of the soft X-ray detections (ROSAT:
    yellow cross; \xmm~: green cross; \sw-XRT: blue cross) are very
    close to each other and practically coincident to that of the
    optical counterpart.}
\label{figura08}
\end{figure}

The analysis of the 54-month BAT significance map shows a considerable
hard X-ray emission at this position but no source has been included
in the 2PBC~catalogue (\cite{cusumano2010b}) because its detection
significance ($3.4~\sigma$) is below the catalogue threshold
$\sigma_T$.
In this case, establishing the position of the source is a delicate task; we
report the coordinates of the pixel with the locally higher significance value
($3.5~\sigma$) that are RA=$01^h 37^m 48^s.77$ and Dec=$+58^{\circ} 11'
21''.1$.
An hard X-ray detection in this region of the sky by \igr~has been
first included in the all-sky survey by Krivonos et~al.~(2007).  
It is also reported in the 4th~IBIS/ISGRI catalogue (\cite{bird2010})
with RA=$01^h 37^m 22^s.8$, Dec=$+58^{\circ} 15' 03''.6$ and with an
error radius of $5'$, thus shrinking the region for the search of a
counterpart to the Fermi $\gamma$-ray source.  
The variability of this object in the 20-40~keV band has been recently
assessed by Telezhinsky et~al.~(2010): it is still classified as
``{\it unidentified}'' and as a transient source detected at the
intrinsic variance map but not at the significance map.

At lower energies, a source in the LAT and IBIS error boxes has been reported
in the 1RXS catalogue (\cite{voges1999}) at RA=$01^h 37^m 48^s.0$,
Dec=$+58^{\circ} 14' 22''.5$ with an error radius of $9''$.
Later on, an X-ray observation performed by \xmm~on January 16, 2003 and aimed
at observing the pulsar PSR~0136+57 has revealed a serendipitous source that
has been included in the 2XMMi catalogue (\cite{watson2009}) with coordinates
RA=$01^h 37^m 50^s.4$, Dec=$+58^{\circ} 14' 10''$ and an error radius of
$1''$.
We analyse this \xmm~observation, characterised by a long exposure time
$t_{exp}$ = 8439~s, considering the results obtained by the combined analysis
of MOS1 and MOS2 detectors.
We fit the spectrum with a power-law model and obtain a photon index $\Gamma =
(2.31 \pm 0.06)$.
The value of the hydrogen column density $N_H = (4.96 \pm 0.18) \times 10^{21}$
cm$^{-2}$ that we obtain leaving this parameter free to vary is very similar
to the Galactic one ($N_H = 4.01 \times 10^{21}$ cm$^{-2}$) as reported in the
Leiden/Argentine/Bonn (LAB) Survey (\cite{kalberla2005}).
The source flux is $1.04 \times 10^{-11}$ erg cm$^{-2}$ s$^{-1}$ in the 2--10
keV band and is $1.5 \times 10^{-11}$ erg cm$^{-2}$ s$^{-1}$ in the 0.2--12 keV
band.
We note the lower value of the flux $1.8 \times 10^{-12}$ erg cm$^{-2}$
s$^{-1}$ in the 2--10 keV band reported by Stephen et al.~(2010) and attribute
it to the different value of the photon index $\Gamma = 1.7$ that they adopted
in their fit.

Two \sw~observations have been recently obtained in this region: the XRT
exposure of the first observation (September 4, 2010) is 1153~s, while a longer
exposure of 3393~s is available for the second observation (October 22, 2010).
Across this period the source has not shown appreciable variations of
activity, with a stable count rate around $\sim 1.7 \times 10^{-1}$
cts~s$^{-1}$.
The position of the XRT source is RA=$01^h 37^m 50^s.37$ and Dec=$+58^{\circ}
14' 11''.5$ with an error radius $3.61'$.
We fit the obtained spectrum with a power-law model and find
$\chi_r^2$ / d.o.f. = 0.96 / 35.
As for the \xmm~observation, also in this case the obtained value for
the hydrogen column density $N_H = (4.75 \pm 0.49) \times 10^{21}$
cm$^{-2}$ can be considered consistent with the Galactic one.
The value of the photon spectral index is $\Gamma = (2.21 \pm 0.14)$, while
the obtained value for the 2--10~keV flux is $5.67 \times 10^{-12}$ erg
cm$^{-2}$ s$^{-1}$ that is nearly half the value measured by \xmm~in
2003.

Possible interesting radio counterparts are in the 87GB and NVSS
(\cite{condon1998}) catalogues (RA=$01^h 37^m 50^s.46$; Dec=$+58^{\circ} 14'
11''.2$) with flux densities of $F_{5~GHz} = 136$~mJy and $F_{1.4~GHz} =
170$~mJy, respectively, from which we derive a spectral index $\alpha_r~\simeq
0.28$.
The image in the latter survey shows a marginally extended source with
a very compact core.

Optical observations (\cite{bikmaev2008}) aimed at the identification
of five \igr~sources reported by Krivonos et~al.~(2007), performed
with the Russian-Turkish 1.5-m RTT-150 and the 6-m BTA telescopes, led
to the discovery of an object having a continuum without emission or
absorption lines, located at RA=$01^h 37^m 50^s.45$ and
Dec=$+58^{\circ} 14' 11''.6$, fully compatible with the radio and
X-ray position and at a separation of $\sim 0.8'$ from the
$\gamma$-ray centroid.
Photometric data of the object in the field are available from the Sloan
Digital Sky Survey (SDSS-DR8).
The morphological classification is that of a starlike source and the
magnitudes are $r = (18.12 \pm 0.01)$~mag, $g = (19.05 \pm 0.01)$~mag and $u =
(20.04 \pm 0.05)$~mag, with a reddening in the $r$~band equal to 1.47~mag.
This implies a correction to the $u-r$ colour index of $\sim 1.2$~mag and the
intrinsic colour index would be $\sim$~0.7~mag, corresponding to a spectral
distribution with a strong excess at blue wavelengths, typical of BL~Lac
objects. 

\begin{figure}[htbp]
\begin{center}
\includegraphics[width=6.9cm, angle=-90]{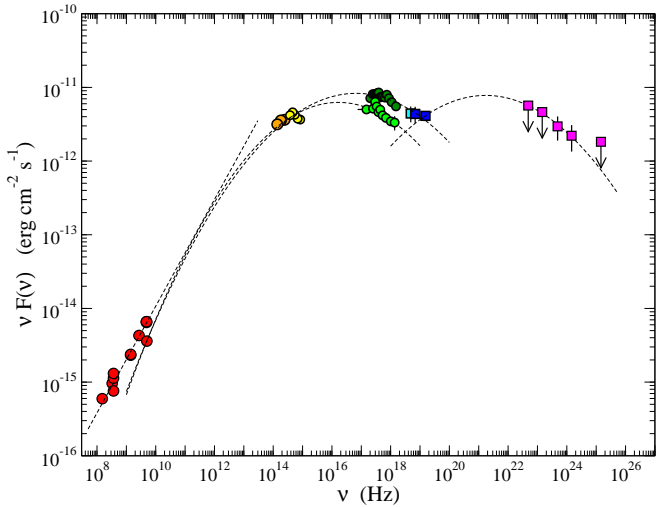}
\end{center}
\caption{The Spectral energy distribution of \1u. From lower to higher
  frequencies we report radio data up to 5 GHz (red circles) and data
  from 2MASS (orange circles), SDSS-DR8 (yellow circles), \sw-XRT
  (light green circles), XMM-MOS (dark green circles), \sw-BAT (cyan
  square), Integral-IBIS (blue squares) and \frm-LAT (magenta
  squares). Data are fitted with a power-law model in the radio band;
  a log-parabolic model has been used to emphasise the synchrotron and
  inverse Compton component of this high-energy synchrotron peak (HSP)
  BL~Lac object.}
\label{figura05}
\end{figure}

Considering all these available data we obtain the spectral energy
distribution of this source shown in Fig.~\ref{figura05}.
It looks very similar to a SED for a LAT blazar (\cite{abdo2010c}) with the
two broad bumps associated with the synchrotron and inverse Compton
emission.
We estimate the peak frequency $\nu_p$ and the curvature parameter $\beta$ of
both components fitting a log-parabolic law to the data.
The observations of the X-ray telescopes, monitoring the region close to the
synchrotron peak in two different epochs, have shown a relevant variability of
emission from this source.
The results of the log-parabolic fit including \sw~data are $\nu_S=1.66 \times
10^{+16}$~Hz for the peak frequency with a corresponding maximum of
synchrotron emission $\nu_S$~F($\nu_S$) = $6.25 \times 10^{-12}$ erg cm$^{-2}$
s$^{-1}$; the estimated value of the curvature parameter is $\beta_{S1} =
0.08$.
The log-parabolic fit including \xmm~data provides a higher value ($\nu_S=6.69
\times 10^{+16}$~Hz) for the synchrotron peak frequency and a corresponding
maximum of synchrotron emission $\nu_S$~F($\nu_S$) = $8.30 \times 10^{-12}$
erg~cm$^{-2}$~s$^{-1}$; a very similar curvature ($\beta_{S2} = 0.07$) was
found also in this case.
As regards the inverse Compton component, we obtain $\nu_{IC}=1.78 \times
10^{+21}$~Hz with a corresponding $\nu_{IC}$~F($\nu_{IC}$) = $7.78 \times
10^{-12}$ erg~cm$^{-2}$~s$^{-1}$; the curvature parameter is $\beta_{IC} =
0.07$.
Both the flux at the peak frequency and the curvature parameter that
we derive from our log-parabolic models are therefore very similar
for the synchrotron and the inverse Compton components.

A BL~Lac classification for this source, first suggested by Bikmaev
et~al.~(2008), has been recently confirmed by Stephen et~al.~(2010).
Following the classification scheme reported in Abdo et~al.~(2010c) we can
conclude from our analysis that this object is an high-energy synchrotron peak
(HSP) BL~Lac object with a $\nu_S > 10^{+16}$~Hz in different detected states
of source's activity.

\subsection{\2u}
\label{sub02}

The position of \2u (RA=$20^h 56^m 43^s.51$, Dec=$+49^{\circ} 38' 37''.3$,
corresponding to the Galactic latitude $b=2^{\circ}.74$), can be assumed with
an error radius $r=2'.7$ from the 1FGL catalogue.

In the 2PBC there is the source 2PBC~J2056.5+4938 close to this
location at RA=$20^h 56^m 32^s.66$, Dec=$+49^{\circ} 38' 30''.3$ with
an error radius $\sim 3'.5$ at 95\% confidence level.
The significance of the detection is 5.5~$\sigma$ in the 15--150 keV
map and 6.3~$\sigma$ in the 15--30 keV map.
Earlier than \sw-BAT, a detection in the hard X-ray band in the
proximity of \2u was obtained by \igr-IBIS.
The source, first reported by Krivonos et~al.~(2007) with the name
IGR~J20569-4940, was left without classification; the detection was
later confirmed by Bird et~al.~(2010) in their 4th IBIS/ISGRI soft
gamma-ray survey catalogue.

The counterpart of this high-energy emission can be searched by
shrinking the region of the sky with circles centered at the positions
of \frm-LAT, \sw-BAT and \igr-IBIS centroids with a radius
proportional to their errors, respectively.
The radio source 4C~+49.35, classified as symmetric double in NED, is found
within the intersection of these circles.
Radio measurements at 1.4~GHz reported in the NVSS catalogue (\cite{condon1998})
resolve two components separated by $\sim3'$: a North-East component
(RA=$20^h 56^m 42^s.69$, Dec=$+49^{\circ} 40' 05''.6$) with a flux density
$F_{NE}=167$~mJy and a South-West component (RA=$20^h 56^m 29^s.45$,
Dec=$+49^{\circ} 38' 01''.0$) with $F_{SW}=124$~mJy.

A soft X-ray emission was detected in this region for the first time
by the Ariel~V satellite and reported by Warwick et~al.~(1981) with
the name 3A~2056+493.
Later detections by ROSAT and, more recently, by \sw~ and \xmm~are
characterised by adequate precision to associate the soft X-ray
emission with the NE component.
Two detections from \xmm~have been reported in the slew survey clean
source catalogue (\cite{saxton2008}).
Their positions are very similar and separated by $\sim 2.5''$ from
each other: RA=$20^h 56^m 42^s.84$, Dec=$+49^{\circ} 40' 03''.8$ for
the first one and RA=$20^h 56^m 42^s.59$; Dec=$+49^{\circ} 40' 04''.3$
for the second one; the error on the position is $8''$ at 1~$\sigma$
confidence.
Despite the two slews were carried out along the same day (November
03, 2007) flux variations in the 0.2--12 keV band have been reported
in the catalogue, dropping from ($1.85 \pm 0.31$) to ($0.83 \pm 0.25$)
$\times 10^{-11}$ erg~cm$^{-2}$~s$^{-1}$ in a few hours.

Two soft X-ray pointed observations have been performed by \sw~on February
26, 2009 and March 03, 2009; the first one has a much longer exposure
(8377~s) than the other (1439~s).
We stack the two observations and obtain RA=$20^h 56^m 42^s.68$;
Dec=$+49^{\circ} 40' 07''.69$ for the position of the X-ray detection,
with a precision of $r=3.54'$.
We carry out a spectral analysis adopting a power-law model: leaving all the
parameters free to vary we obtain $N_H = (1.62 \pm 0.09) \times 10^{22}$
cm$^{-2}$, $\Gamma = (2.43 \pm 0.08)$ and $\chi_r^2$ / d.o.f. = 0.94 / 97; the
2--10 keV flux is $1.19 \times 10^{-11}$ erg cm$^{-2}$ s$^{-1}$.
\begin{figure}[htbp]
\begin{center}
\includegraphics[width=6.9cm, angle=-90]{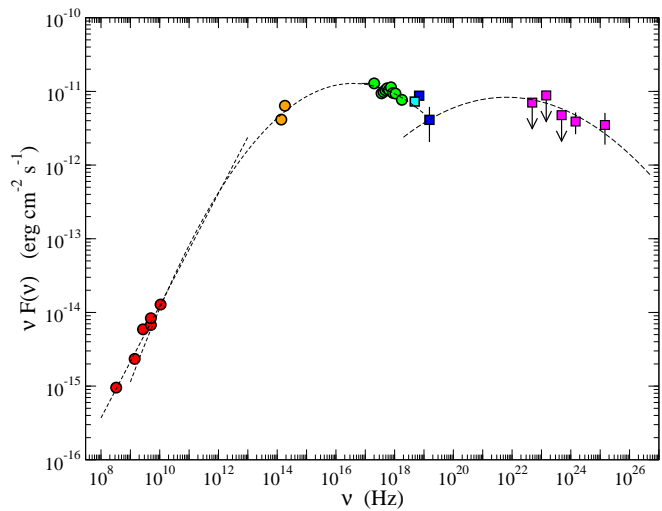}
\end{center}
\caption{The Spectral energy distribution of \2u. From lower to higher
  frequencies we report radio data up to 22~GHz (red circles) and data
  from 2MASS (orange circles), \sw-XRT (light green circles), \sw-BAT
  (cyan square), Integral-IBIS (blue squares) and \frm-LAT (magenta
  squares). Data are fitted with a power-law model in the radio
    band; a log-parabolic model has been used to emphasise the
    synchrotron and inverse Compton component of this high-energy
    synchrotron peak (HSP) BL~Lac object.}
\label{figura06}
\end{figure}
We note that, at variance with our result, Landi et~al.~(2010)
report a value of the hydrogen column density $N_H =(0.53 \pm 0.18)
\times 10^{22}$ cm$^{-2}$ lower than the Galactic contribution ($N_H =
1.0 \times 10^{22}$ cm$^{-2}$) quoted in the LAB Survey
(\cite{kalberla2005}).
We repeat the fit fixing the $N_H$ parameter to the Galactic one but we find a
not acceptable variation in the $\chi^2$ value ($\chi_r^2$ / d.o.f. = 1.61 / 96).
A broad band spectral fit of data from radio to the X-rays with a log-parabola
provides a curvature parameter $\beta_S=0.07$ (Fig.~\ref{figura06}).
Therefore we repeat the fit of the X-ray spectrum with a log-parabolic model
fixing $\beta$ = 0.07 and also the hydrogen column density at a value ($N_H =
1.2 \times 10^{22}$ cm$^{-2}$) moderately higher with respect to the Galactic
one: in this way we obtain the acceptable result \- $\chi_r^2$ / d.o.f. = 1.14
/ 94.

The association between the radio and the soft X-ray emission can be
considered firmly established by the short distance between the
\sw~and the \xmm~centroids on the one hand and the NE~component of
4C~+49.35 on the other hand.
These are very close to the bright star SAO~50269 (RA=$20^h 56^m
41^s.966$; Dec=$+49^{\circ} 40' 17''.55$; $R_2=8.39$ mag in USNO-B1,
\cite{monet2003}).
The UVOT image in the UV-W2 filter does not show evidence of sources
in proximity of the star, for which we obtain a magnitude
$W2 = (13.12 \pm 0.05)$~mag.  
Both images acquired in the U~filter are saturated in correspondence
to the star.
In a study of the low latitude sample area in Cygnus of the ROSAT
galactic plane survey, Motch et~al.~(1997) associated the X-ray source
RX~J2056.6+4940 with SAO~50269 with a probability for a random
association of about 10\%.
The distance between the X-ray and the star position is $\sim 13''$ within the
estimated ROSAT accuracy that is $\sim 25''$.
The origin of the X-ray emission was related to an active corona surrounding 
this star, as for the 85\% of sources with a PSPC count rate higher than 
$3 \times 10^{-2}$ cts~s$^{-1}$ belonging to their sample.
Recently, Haakonsen \& Rutledge (2009) reported a low probability of
association between 1RXS~J205644.3+494011 and the star in their statistical
cross-association of the ROSAT Bright Source Catalogue (\cite{voges1999}) and
the 2MASS Point Source Catalogue (\cite{cutri2003}).
As reported by Landi et~al.~(2010) the counterpart of this high-energy
source is, with high probability, an object located at RA=$20^h 56^m
42^s.719$ and Dec=$+49^{\circ} 40' 06''.9$ at only $12''.9$ from
SAO~50269.
It has been detected in the 2MASS infrared images retrieved from the
NASA/IPAC infrared Science Archive.
The magnitude in the J~filter is 13.69~mag and is uncertain, whereas
in the H and K filters they are $(14.30 \pm 0.10)$~mag and $(13.74 \pm
0.08)$~mag, respectively.
The emission of this object in the optical as well as in the UV band
is most likely overwhelmed by that originating from the star.
This source is separated from the $\gamma$-ray centroid by $1'.5$.

Paredes et~al.~(2002) cited the ROSAT source 1RXS~J205644.3+494011 in
a search for microquasar candidates at low galactic latitudes obtained
by means of a cross-identification between the ROSAT bright source
catalogue (RBSC, \cite{voges1999}) and the NVSS catalogue.
These authors gave a priority to sources for which the offset between
the X-ray and the radio position was within the 1$\sigma$ RBSC
position error: unfortunately, this was not the case for
1RXS~J205644.3+494011.
Recently, the hypothesis that the nature of this source might be
extragalactic was taken into account by several authors (Landi et
al.~(2010); Voss\&Ajello~(2010); Stephen et~al.~(2010)).
We collect for the first time all the available data in literature
to build the SED of this object, that is reported in
Fig.~\ref{figura06}, to investigate the broad-band emission properties
of \2u.

We exclude the radio measurements reported by surveys with a low
resolution to avoid the risk of including a spurious contribution from
the SW component of 4C~+49.35.
We estimate a radio spectral index $\alpha_r~\sim~0.24$ from the fit
of data up to 22~GHz (\cite{petrov2007}): this value of $\alpha_r$ is
indeed typical of flat spectrum radio sources.
An evaluation of correct fluxes both at optical and at infrared
wavelengths is difficult not only for the near star but also for the
high value of the reddening $E(B-V)=2.85$~mag (\cite{schlegel1998}),
that may be uncertain for such a low value of the Galactic latitude.
We fit with a log-parabola the data from radio to X-ray frequencies and obtain
the peak frequency $\nu_S=3.86 \times 10^{+16}$~Hz, the corresponding maximum
of synchrotron emission $\nu_S$~F($\nu_S$) = $1.28 \times 10^{-11}$
erg~cm$^{-2}$~s$^{-1}$ and the curvature parameter $\beta_{S} = 0.07$.
In carrying out our fit we force the log-parabola profile to agree
with the power-law fit to the radio data and assume $E(B-V)=1$~mag,
consistently lower than the value quoted by Schlegel et~al.~(1998);
the uncertain infrared measurement in the J~filter has been omitted.
We also carry out a fit of high-energy data from the hard X- to the
$\gamma$-ray band and obtain $\nu_{IC}=6.46 \times 10^{+21}$~Hz with a
corresponding $\nu_{IC}$~F($\nu_{IC}$) = $8.32 \times 10^{-12}$
erg~cm$^{-2}$~s$^{-1}$ and a curvature parameter $\beta_{IC} = 0.04$.

More precise measurements are requested to characterise both emission
components of this source.
Nevertheless, the analysis of its broad-band properties such as the
radio spectral index, the low Compton dominance and the synchrotron
peak frequency higher than $10^{16}$~Hz support for \2u the
classification as a blazar and in particular as a HSP BL~Lac object,
but only the analysis of the optical spectrum can confirm this
interpretation.

\section{Conclusions}
\label{sect:6}

We have reported the results of our analysis devoted to the research of the
sources showing high-energy emission in both the \sw-BAT and the \frm-LAT
telescopes, according to the data collected by LAT across the first 11 months
of operation and by BAT across a much longer period of 54 months.
As expected, we have found a low number of sources: only 7\% of those included
in the 1FGL catalogue are characterised by a significant emission ($\sigma >
3$) as shown by the 54-month BAT 15-150 keV map.
The larger fraction of them is given by extragalactic objects, and the
dominant part is represented by blazars. 
We have investigated with greater detail their redshift distribution making a
distinction among its subclasses: the comparison with the distribution of the
whole population classified in the second edition of the BZCAT catalogue shows
that they are mainly closer than the average to the observer.

Driven by the detection from both the BAT and the LAT instruments we have
focused on a couple of objects with a very low Galactic latitude.
Our detailed analysis of their broad-band spectral energy distribution
supports the classification of these objects as blazars, and in particular as
high synchrotron peaked (HSP) BL~Lac Objects.
Analyses of the same kind can be carried out with success on other sources
that are expected in the forthcoming future with the release of the second
Fermi-LAT point source catalogue, based on two years of $\gamma$-ray data, and
also a new Palermo BAT catalogue obtained from the analysis of six years of
BAT data.
    
\begin{acknowledgements}
  The authors are grateful to the referee for the suggestions and
  comments that helped to improve the manuscript. They acknowledge
  financial support by ASI/INAF through contract I/011/07/0. Part of
  this work is based on archival data, software or on-line services
  provided by the ASI Science Data Center (ASDC) and by the SIMBAD
  database which is operated at CDS, Strasbourg, France.
\end{acknowledgements}

\setcounter{table}{0}
\begin{table*}[thbp]
\vspace{0.5cm}
\caption{The list of correspondences between the 1FGL and 2PBC
  catalogues at galactic latitude $|b| > 10^{\circ}$. For each high
  energy source the counterpart associated in the corresponding
  catalogues, and the relative classification, is reported: their
  agreement is marked by a code in the last column. In a few cases
  marked with an asterisk ($^\ast$) the counterpart of the 1FGL source
  comes from Abdo et~al.~(2010b). The parameter $Q$ gives an estimate
  of the reliability of the association based on the overlap between
  the error circles: it lacks for 1FGL sources with no error
  estimate.}
\vspace{0.5cm}
\label{tabella02}
\begin{center}
\begin{tabular}{l|c|c||l|c|c||c|c}
1FGL source          & 1FGL counterpart               & type & 2PBC source         & 2PBC counterpart        & type & $Q$  & code \\ 
\hline                                                                                                       
1FGL~J0101.3$-$7257  & SMC                            & gal  & 2PBC~J0102.7$-$7241 & XTE J0103$-$728         & HXB  & 0.86 & n    \\
                     &                                &      &                     & RX J0104.1$-$7244       & HXB  & 0.86 & n    \\  
1FGL~J0217.8+7353    & 1ES 0212+735                   & bzq  & 2PBC~J0217.4+7349   & 1ES 0212+735            & BZQ  & 0.47 & y    \\
1FGL~J0238.3$-$6132  & PKS 0235$-$618                 & bzq  & 2PBC~J0238.3$-$6116 & IRAS F02374$-$6130      &   G  & 0.94 & n    \\
1FGL~J0242.7+0007    & RXJ0241.9+0009~($^{\ast}$)     & unk  & 2PBC~J0242.7$-$0000 & NGC 1068                & Sy2  & 0.88 & n    \\           
1FGL~J0319.7+4130    & NGC 1275                       & agn  & 2PBC~J0319.7+4129   & NGC 1275                & BZU  & 0.41 & y    \\
1FGL~J0325.0+3403    & B2 0321+33B                    & agn  & 2PBC~J0324.7+3409   & 1H 0323+342             & Sy1  & 0.46 & y    \\
1FGL~J0334.2+3233    & NRAO 140~($^{\ast}$)           & bzq  & 2PBC~J0336.5+3219   & NRAO 140                & BZQ  & 1.27:& y    \\
1FGL~J0339.1$-$1734  & PKS 0336$-$177                 & agn  & 2PBC~J0339.2$-$1742 & 1RXS J033913.4$-$173553 &   X  & 1.49 & y    \\             
1FGL~J0405.6$-$1309  & PKS 0403$-$13                  & bzq  & 2PBC~J0405.6$-$1308 & RX J0405.5$-$1308       & BZQ  & 0.43 & y    \\
1FGL~J0437.2$-$4715  & PSR J0437$-$4715               & PSR  & 2PBC~J0437.8$-$4713 & RBS 0560                & Sy1  & ---  & n    \\             
1FGL~J0440.6+2748    & B2 0437+27B                    & bzb  & 2PBC~J0440.8+2739   & 1RXS J044046.9+273948   &   X  & 1.38:& n    \\
1FGL~J0522.8$-$3632  & PKS 0521$-$36                  & bzb  & 2PBC~J0523.0$-$3626 & RBS 0644                & BZU  & 1.01 & y    \\
1FGL~J0531.0+1331    & PKS 0528+134                   & bzq  & 2PBC~J0530.9+1333   & PKS 0528+134            & BZQ  & 0.52 & y    \\
1FGL~J0538.8$-$4404  & PKS 0537$-$441                 & bzb  & 2PBC~J0538.9$-$4406 & PKS 0537$-$441          & BZB  & 0.46 & y    \\
1FGL~J0539.1$-$2847  & PKS 0537$-$286                 & bzq  & 2PBC~J0539.8$-$2839 & PKS 0537$-$286          & BZQ  & 0.56 & y    \\
1FGL~J0557.6$-$3831  & CRATES J0558$-$3838            & bzb  & 2PBC~J0558.0$-$3821 & H 0557$-$385            & Sy1  & 1.21 & n    \\
1FGL~J0636.1$-$7521  & PKS 0637$-$75                  & bzq  & 2PBC~J0635.4$-$7514 & PKS 0637$-$752          & BZQ  & 0.80 & y    \\
1FGL~J0710.6+5911    & BZB J0710+5908                 & bzb  & 2PBC~J0710.2+5909   & 1H 0658+595             & BZB  & 0.58 & y    \\
1FGL~J0746.6+2548    & B2 0743+25                     & bzq  & 2PBC~J0746.4+2548   & 87GB 074322.5+255639    & BZQ  & 0.31 & y    \\
1FGL~J0750.6+1235    & PKS 0748+126                   & bzq  & 2PBC~J0750.6+1231   & OI +280                 & BZQ  & 0.43 & y    \\
1FGL~J0806.2+6148    & CGRaBS J0805+6144              & bzq  & 2PBC~J0805.4+6146   & GB6 J0805+6144          & BZQ  & 0.87 & y    \\
1FGL~J0842.2+7054    & 4C +71.07                      & bzq  & 2PBC~J0841.4+7053   & S5 0836+71              & BZQ  & 0.37 & y    \\
1FGL~J0929.4+5000    & CGRaBS J0929+5013~($^{\ast}$)  & bzb  & 2PBC~J0928.5+4959   & ---                     & ---  & 0.96 & -    \\
                     &                                &      & 2PBC~J0930.6+4954   & RBS 0782                & BZB  & 1.29:& n    \\
1FGL~J0949.0+0021    & CGRaBS J0948+0022              & agn  & 2PBC~J0948.9+0021   & RX J0948.8+0022         & BZQ  & 0.27 & y    \\
1FGL~J0956.5+6938    & M 82                           & sbg  & 2PBC~J0955.7+6941   & M 82                    &  IG  & 0.64 & y    \\             
1FGL~J1048.7+8054    & CGRaBS J1044+8054              & bzq  & 2PBC~J1044.1+8054   & S5 1039+81              & BZQ  & 1.07 & y    \\
1FGL~J1103.7$-$2329  & CRATES J1103$-$2329            & bzb  & 2PBC~J1103.6$-$2329 & 1H 1100$-$230           & BZB  & 0.37 & y    \\
1FGL~J1104.4+3812    & Mkn 421                        & bzb  & 2PBC~J1104.4+3813   & Mrk 421                 & BZB  & 0.41 & y    \\
1FGL~J1130.2$-$1447  & PKS 1127$-$14                  & bzq  & 2PBC~J1130.1$-$1449 & OM $-$146               & BZQ  & 0.37 & y    \\
1FGL~J1136.2+6739    & BZB J1136+6737                 & bzb  & 2PBC~J1137.2+6735   & RBS 1004                & BZB  & 0.18 & y    \\
1FGL~J1221.3+3008    & B2 1218+30                     & bzb  & 2PBC~J1221.3+3008   & 1RXS J122121.7+301041   & BZB  & 0.12 & y    \\
1FGL~J1222.5+0415    & 4C +04.42                      & bzq  & 2PBC~J1222.3+0415   & 4C 04.42                & BZQ  & 0.62 & y    \\
1FGL~J1224.7+2121    & 4C +21.35                      & bzq  & 2PBC~J1224.8+2122   & 4C +21.35               & BZQ  & 0.40 & y    \\
1FGL~J1227.9$-$4852  & ---                            & ---  & 2PBC~J1228.0$-$4854 & XSS J12270$-$4859       & CV*  & 0.22 & -    \\       
1FGL~J1229.1+0203    & 3C 273                         & bzq  & 2PBC~J1229.1+0202   & 3C 273                  & BZQ  & 0.25 & y    \\
1FGL~J1256.2$-$0547  & 3C 279                         & BZQ  & 2PBC~J1256.1$-$0547 & 3C 279                  & BZQ  & 0.37 & y    \\
1FGL~J1305.4$-$4928  & NGC 4945                       & agn  & 2PBC~J1305.4$-$4928 & NGC 4945                & Sy2  & 0.10 & y    \\           
1FGL~J1307.0$-$4030  & ESO 323$-$77~($^{\ast}$)       & agn  & 2PBC~J1306.5$-$4025 & ESO 323$-$77            & Sy2  & 0.74 & y    \\            
1FGL~J1320.1$-$4007  & ---                            & ---  & 2PBC~J1320.2$-$4014 & ---                     & ---  & 0.72 & -    \\
1FGL~J1325.6$-$4300  & Cen A                          & agn  & 2PBC~J1325.4$-$4301 & Cen A                   & BZU  & 0.61 & y    \\
1FGL~J1331.9$-$0506  & PKS 1329$-$049                 & bzq  & 2PBC~J1332.0$-$0510 & PKS 1329$-$049          & BZQ  & 0.91 & y    \\
1FGL~J1417.8+2541    & 2E 1415+2557                   & bzb  & 2PBC~J1417.9+2543   & 7C 1415+2556            & BZB  & 0.37 & y    \\
1FGL~J1428.7+4239    & 1ES 1426+428                   & bzb  & 2PBC~J1428.6+4239   & H 1426+428              & BZB  & 0.58 & y    \\
1FGL~J1442.8+1158    & 1ES 1440+122                   & bzb  & 2PBC~J1442.8+1202   & RBS 1420                & BZB  & 0.90 & y    \\
1FGL~J1512.8$-$0906  & PKS 1510$-$08                  & BZQ  & 2PBC~J1512.8$-$0906 & PKS 1510$-$08           & BZQ  & 0.42 & y    \\
1FGL~J1517.8$-$2423  & AP Lib                         & bzb  & 2PBC~J1517.7$-$2419 & Ap Lib                  & BZB  & 0.20 & y    \\
1FGL~J1555.7+1111    & PG 1553+113                    & bzb  & 2PBC~J1555.5+1109   & PG 1553+113             & BZB  & 0.73 & y    \\
1FGL~J1626.2$-$2956  & PKS 1622$-$29                  & bzq  & 2PBC~J1626.0$-$2952 & PKS 1622$-$29           & BZU  & 0.85 & y    \\
1FGL~J1642.5+3947    & 3C 345~($^{\ast}$)             & bzq  & 2PBC~J1643.0+3951   & 4C 39.48                & BZQ  & 1.52 & y    \\
\hline
\multicolumn{8}{c} { }                                                                                                     
\end{tabular}
\end{center}
\end{table*}

\setcounter{table}{0}
\begin{table*}[thbp]
\vspace{0.5cm}
\caption{continues} 
\vspace{0.5cm}
\begin{center}
\begin{tabular}{l|c|c||l|c|c||c|c}
1FGL source          & 1FGL counterpart & type & 2PBC source         & 2PBC counterpart & type & $Q$  & code \\ 
\hline                                                                                                
1FGL~J1653.9+3945    & Mkn 501          & bzb  & 2PBC~J1653.8+3945   & Mrk 501          & BZB  & 1.16 &  y   \\
1FGL~J1829.8+4845    & 3C 380           & agn  & 2PBC~J1829.6+4845   & 3C 380           & BZU  & 0.24 &  y   \\
1FGL~J1835.3$-$3255  & NGC 6652         & glc  & 2PBC~J1835.7$-$3259 & XB 1832$-$330    & LXB  & 0.80 &  n   \\      
1FGL~J1925.2$-$2919  & PKS B1921$-$293  & bzq  & 2PBC~J1924.4$-$2913 & OV $-$236        & BZQ  & 1.47 &  y   \\
1FGL~J2000.0+6508    & 1ES 1959+650     & bzb  & 2PBC~J1959.8+6509   & 1ES 1959+650     & BZB  & 1.10 &  y   \\
1FGL~J2011.4$-$2903  & ---              & ---  & 2PBC~J2010.8$-$2910 & ---              & ---  & 1.71 &  -   \\
1FGL~J2148.5+0654    & 4C +06.69        & bzq  & 2PBC~J2148.0+0657   & 4C +06.69        & BZQ  & 0.41 &  y   \\
1FGL~J2202.8+4216    & BL Lac           & bzb  & 2PBC~J2202.7+4217   & BL Lac           & BZB  & 0.24 &  y   \\
1FGL~J2229.7$-$0832  & PKS 2227$-$08    & bzq  & 2PBC~J2229.6$-$0831 & PKS 2227$-$08    & BZQ  & 0.65 &  y   \\
1FGL~J2232.5+1144    & CTA 102          & bzq  & 2PBC~J2232.4+1144   & 4C +11.69        & BZQ  & 0.11 &  y   \\
1FGL~J2253.9+1608    & 3C 454.3         & BZQ  & 2PBC~J2253.9+1609   & 3C 454.3         & BZQ  & 0.46 &  y   \\
1FGL~J2327.7+0943    & PKS 2325+093     & bzq  & 2PBC~J2327.4+0939   & PKS J2327+0940   & BZQ  & 1.22 &  y   \\
1FGL~J2359.0$-$3035  & 1H 2351$-$315    & bzb  & 2PBC~J2359.1$-$3035 & H 2356$-$309     & BZB  & 0.28 &  y   \\
\hline
\multicolumn{8}{c} { }
\end{tabular}
\end{center}
\end{table*}

\begin{table*}[thbp]
\vspace{2.5cm}
\caption{The list of correspondences between the 1FGL and 2PBC catalogues at
  galactic latitude $|b| < 10^{\circ}$.}
\label{tabella03}
\vspace{0.5cm}
\begin{center}
\begin{tabular}{l|c|c||l|c|c||c|c}
1FGL source          & 1FGL counterpart & type & 2PBC source         & 2PBC counterpart             & type & $Q$  & code \\
\hline                                                                                                           
1FGL~J0035.9+5951    & 1ES~0033+595     & bzb  & 2PBC~J0035.8+5951   & 1ES 0033+59.5                & BZB  & 0.49 & y    \\
1FGL~J0240.5+6113    & LS I+61 303      & HXB  & 2PBC~J0240.6+6114   & GT 0236+610                  & HXB  & ---  & y    \\
1FGL~J0419.0+3811    & 3C 111           & agn  & 2PBC~J0418.3+3801   & 3C 111                       & Sy1  & 0.86 & y    \\
1FGL~J0534.5+2200    & PSR J0534+2200   & PSR  & 2PBC~J0534.5+2201   & Crab                         & PSR  & ---  & y    \\
1FGL~J0730.3$-$1141  & PKS 0727$-$11    & bzq  & 2PBC~J0730.4$-$1142 & PG 0727$-$11                 & BZQ  & 0.37 & y    \\
1FGL~J0835.3$-$4510  & PSR J0835$-$4510 & PSR  & 2PBC~J0835.3$-$4511 & Vela Pulsar                  & PSR  & ---  & y    \\
1FGL~J1045.2$-$5942  & ---              & ---  & 2PBC~J1044.8$-$5942 & V* eta Car                   &  V*  & 0.70 & -    \\
1FGL~J1124.6$-$5916  & PSR J1124$-$5916 & PSR  & 2PBC~J1124.9$-$5919 & SNR G292.0+01.8              & SNR  & ---  & y    \\
1FGL~J1632.7$-$4733c & ---              & ---  & 2PBC~J1632.7$-$4727 & IGR J16328$-$4726            & gam  & 0.87 & -    \\
1FGL~J1656.2$-$3257  & ---              & ---  & 2PBC~J1656.2$-$3303 & SWIFT J1656.3$-$3302         & QSO  & 0.85 & -    \\
1FGL~J1724.0$-$3611c & ---              & ---  & 2PBC~J1725.2$-$3616 & IGR J17252$-$3616            & HXB  & 0.69:& -    \\
1FGL~J1738.5$-$2656  & ---              & ---  & 2PBC~J1738.2$-$2700 & SLX 1735$-$269               & LXB  & 0.74 & -    \\
1FGL~J1746.4$-$2849c & PWN G0.13$-$0.11 & pwn  & 2PBC~J1746.8$-$2845 & CXOGCS J174621.05$-$284343.2 &   X  & 1.00c& n    \\
1FGL~J1747.2$-$2958  & PSR J1747$-$2958 & PSR  & 2PBC~J1747.4$-$3000 & 1RXS J174726.8$-$300008      & LXB  & ---  & n    \\
1FGL~J1747.6$-$2820c & ---              & ---  & 2PBC~J1747.6$-$2820 & CXOGCS J174742.4$-$282228    &   X  & 0.94 & -    \\
1FGL~J1826.2$-$1450  & LS 5039          & HXB  & 2PBC~J1826.3$-$1450 & V* V479 Sct                  & HXB  & ---  & y    \\
1FGL~J1833.5$-$1034  & PSR J1833$-$1034 & PSR  & 2PBC~J1833.5$-$1033 & SNR 021.5$-$00.9             & PSR  & ---  & y    \\
1FGL~J1833.6$-$2103  & PKS 1830$-$21    & bzq  & 2PBC~J1833.7$-$2102 & PKS 1830$-$211               & BZQ  & 0.34 & y    \\
1FGL~J2015.7+3708    & ---              & ---  & 2PBC~J2015.9+3712   & RX J2015.6+3711              & CV*  & 0.95 & -    \\
1FGL~J2032.4+4057    & Cyg X-3          & MQO  & 2PBC~J2032.4+4057   & Cyg X-3                      & HXB  & ---  & y    \\
1FGL~J2056.7+4938    & ---              & ---  & 2PBC~J2056.5+4938   & RX J2056.6+4940              &   X  & 0.43 & -    \\
1FGL~J2323.4+5849    & ---              & spp  & 2PBC~J2323.3+5849   & Cas A                        & SNR  & 0.21 & -    \\
\hline                                                                                                                     
\multicolumn{8}{c} { }                                                                                                     
\end{tabular}                                                                                                       
\end{center}                                                                                                               
\end{table*}

\begin{table*}[thbp]                                                                                                       
  \caption{The list of 1FGL sources at Galactic latitude $|b| >
    10^{\circ}$ obtained considering a significance threshold
    $\sigma_T^{\star} = 3$ in the BAT 15--150 keV all-sky map. For each
    source the association and the classification type indicated in
    the 1FGL~catalogue is reported, together with BZCAT blazar
    classification.}
\label{tabella04}
\begin{center}                                                                                                             
\begin{tabular}{l|c|c|l}                                                                                            
1FGL source         & 1FGL counterpart         & type & BZCAT source     \\  
\hline                                                                                                           
1FGL~J0137.5$-$2428 & PKS 0135$-$247           & bzq  & BZQ~J0137$-$2430 \\  
1FGL~J0144.9$-$2732 & PKS 0142$-$278           & bzq  & BZQ~J0145$-$2733 \\  
1FGL~J0334.2$-$4010 & PKS 0332$-$403           & bzb  & BZB~J0334$-$4008 \\  
1FGL~J0403.9$-$3603 & PKS 0402$-$362           & bzq  & BZQ~J0403$-$3605 \\  
1FGL~J0455.6$-$4618 & PKS 0454$-$46            & bzq  & BZQ~J0455$-$4615 \\  
1FGL~J0507.3$-$6103 & CRATES J0507$-$6104      & bzq  & BZQ~J0507$-$6104 \\  
1FGL~J0507.9+6738   & 1ES 0502+675             & bzb  & BZB~J0507+6737   \\  
1FGL~J0639.9+7325   & CGRaBS J0639+7324        & bzq  & BZQ~J0639+7324   \\  
1FGL~J0721.9+7120   & S5 0716+714              & bzb  & BZB~J0721+7120   \\  
1FGL~J0739.1+0138   & PKS 0736+01              & bzq  & BZQ~J0739+0137   \\  
1FGL~J1031.0+5051   & 1ES 1028+511             & bzb  & BZB~J1031+5053   \\  
1FGL~J1159.4+2914   & 4C +29.45                & bzq  & BZQ~J1159+2914   \\  
1FGL~J1354.9$-$1041 & PKS 1352$-$104           & bzq  & BZU~J1354$-$1041 \\  
1FGL~J1604.3+5710   & CGRaBS J1604+5714        & bzq  & BZQ~J1604+5714   \\  
1FGL~J1617.9$-$7716 & PKS 1610$-$77            & bzq  & BZQ~J1617$-$7717 \\  
1FGL~J1637.9+4707   & 4C +47.44                & bzq  & BZQ~J1637+4717   \\  
1FGL~J1725.0+1151   & CGRaBS J1725+1152        & bzb  & BZB~J1725+1152   \\  
1FGL~J1800.4+7827   & CGRaBS J1800+7828        & bzb  & BZB~J1800+7828   \\  
1FGL~J1849.3+6705   & CGRaBS J1849+6705        & bzq  & BZQ~J1849+6705   \\  
1FGL~J1923.5$-$2104 & OV $-$235                & bzq  & BZQ~J1923$-$2104 \\  
1FGL~J2143.4+1742   & OX 169                   & bzq  & BZQ~J2143+1743   \\  
1FGL~J2158.8$-$3013 & PKS 2155$-$304           & bzb  & BZB~J2158$-$3013 \\  
\hline                                                                                                               
\multicolumn{4}{c} { }                                                                                               
\end{tabular}                                                                                                        
\end{center}                                                                                                         
\end{table*}                                                                                                       

\begin{table*}[thbp]                              
  \caption{The list of 1FGL sources at Galactic latitude $|b| <
    10^{\circ}$ obtained considering a significance threshold
    $\sigma_T^{\star} = 3$ in the BAT 15-150 keV all-sky map.}
\label{tabella05}
\begin{center}                                                                                                       
\begin{tabular}{l|c|c|c}                                                                                            
1FGL source          & 1FGL counterpart   & type & BZCAT source   \\ 
\hline                                                                                                         
1FGL~J0137.8+5814    & ---                & ---  & ---            \\ 
1FGL~J0849.6$-$3540  & VCS2 J0849$-$3541  & agu  & ---            \\ 
1FGL~J1329.2$-$5605  & PMN J1329$-$5608   & agu  & ---            \\ 
1FGL~J1420.1$-$6048  & PSR J1420$-$6048   & PSR  & ---            \\ 
1FGL~J1714.5$-$3830c & ---                & ---  & ---            \\ 
1FGL~J2021.0+3651    & PSR J2021+3651     & PSR  & ---            \\ 
1FGL~J2347.1+5142    & 1ES 2344+514       & bzb  & BZB~J2347+5142 \\ 
\hline
\multicolumn{4}{c} { }
\end{tabular}
\end{center}
\scriptsize{From Abdo et~al.~(2010a): gal = normal galaxy; agn = non
  blazar active galaxy; sbg = starburst galaxy; bzb = BL~Lac object;
  bzq = flat spectrum radio quasar; agu = active galaxy of uncertain
  type; psr = pulsar; pwn = pulsar wind nebula; spp = potential
  association with a supernova remnant or a pulsar wind nebula; glc =
  globular cluster; mqo = microquasar object; hxb = other X-ray
  binary. Designations in capital letters correspond to firm
  identifications. From Abdo et~al.~(2010b): unk = AGN of unknown
  type. The nomenclature adopted in Cusumano et~al.~(2010b) is derived
  from SIMBAD online services, with the exception of blazars: LXB =
  low mass X-ray binary; HXB = high mass X-ray binary; G = normal
  galaxy; Sy1 = Seyfert~1 galaxy; Sy2 = Seyfert~2 galaxy; X = X-ray
  source; IG = interacting galaxy; CV* = cataclismic variable star;
  PSR = pulsar; V* = variable star; SNR = superNova remnant; gam =
  gamma-ray source; QSO = Quasi Stellar object. The nomenclature of
  the BZCAT (\cite{massaro2009}) has been used for blazars: BZB =
  BL~Lac objects; BZQ = flat spectrum radio quasars; BZU = blazars
  with uncertain classification.}
\end{table*}
\end{document}